\documentstyle[aps,epsfig]{revtex}

\newtheorem{theorem}{Theorem}
\newtheorem{proposition}[theorem]{Proposition}

\newtheorem{rem}[theorem]{Remark}

\begin{document}
\bibliographystyle{abbrv}
\title{Semi-classical study of the Quantum Hall conductivity.}
\author{Fr\'ed\'eric Faure 
   \and Bernard Parisse 
}

\maketitle

\begin{abstract}
 The semi-classical study of the integer Quantum Hall conductivity
 is investigated for electrons in a bi-periodic potential $V(x,y)$.
 The Hall conductivity is due to the tunnelling effect and we concentrate our study to
 potentials having three wells in a periodic cell. A non-zero topological conductivity
 requires special conditions for the positions, and shapes of the wells. The results are
 derived analytically and well confirmed by numerical calculations. 
\end{abstract}

\vspace{1cm}

P.A.C.S. number : 03.65.Sq , 73.50.Jt,  73.40.Hm , 73.20.Dx : electrons in low dimensional structures , 05.45.+b  \\
key words: semi classical analysis, tunnelling effect, quantum Hall conductivity  \\
AMS:81Q20 (WKB)
81S30 (phase space method)
82D20 (solids)
 \\
%\tableofcontents
\vspace{0.5cm}

{\em 
\noindent
Fr\'ed\'eric Faure \\
Laboratoire de Physique Num\'erique des Syst\`emes Complexes,
Universit\'e Joseph Fourier/CNRS, 
BP 166, 
38042 Grenoble Cedex 9, 
France, \\
e-mail: {\tt faure@labs.polycnrs-gre.fr}\\

\vspace{0.5cm}

\noindent Bernard Parisse\\
Institut Fourier, 
CNRS UMR 5582, 
38402 St Martin d'H\`eres C\'edex, 
France \\
e-mail: {\tt parisse@fourier.ujf-grenoble.fr}
}

%\pagebreak

\section{Introduction}
 Some physical phenomenon happen to be expressible from topological properties of specific models. 
The integer Hall conductivity is one of them \cite{thouless1},\cite{iqhe2}. 
In a simple model of non-interacting electrons moving in a two-dimensional periodic potential $V(x,y)$ subject to a uniform perpendicular magnetic field $B_z$ and a low electric field $E_y$, the Hall 
conductivity $\sigma_{xy}$ of a given filled Landau electronic band turns out to be proportional 
to an integer $C$:
\[ \sigma_{xy}=\frac{e^2}{h}C \]
 $C$ is the Chern index of the band,
 describing the topology of its fiber bundle structure \cite{chern6},\cite{bellissard1},\cite{thouless1},\cite{chern4}.\\
For a better understanding of this phenomenon, and to bring out the conditions of possible experimental measures, we investigate in this paper the value of $C$ as a function of the potential $V$. This is done by semi-classical methods, and the tunnelling effect appears to be the fundamental mechanism for a non-zero conductivity.\\
In the limit of high magnetic field $B_z$, the above model is mapped onto 
the well-known Harper model, by means of the Peierls substitution: the 
potential $V$ is considered as a perturbation of the cyclotron motion, and 
the averaging method of mechanics gives an effective Hamiltonian equal to 
the average of $V$ on the cyclotron circles. We neglect the coupling between the Landau bands \cite{geisel1}. For a high magnetic field, (hence for a small cyclotron radius,) this transformation gives an effective Hamiltonian $H_{eff}(q,p)\sim V(q,p)$, which is biperiodic in position and momentum (the phase space is a 2D torus), and an effective Planck constant $h_{eff}=hc/(eB)$. In this approximation, trajectories are the levels lines of 
$H_{eff}\sim V$. Furthermore, the expression of $h_{eff}$ shows that the high magnetic field regime corresponds to the semi-classical limit. This model will be the starting point of our study in the next section.\\
Quantum mechanics on the torus has been extensively used as a convenient 
framework
to study basic properties of the semi-classical limit like quantum chaos 
(\cite{chern2},\cite{zero2},\cite{saraceno90}, \cite{debievre96}), 
or the tunnelling effect (\cite{harper8,harper4}, 
\cite{helffer90}).
For this purpose, a convenient tool is the Bargmann and Husimi representation which maps a quantum state to a function on the phase space.  These representations are constructed with coherent states and will be recalled in section 3.\\
The new results presented in this paper are the conditions under which the tunnelling effect between three different wells can be responsible 
for a non-zero Hall conductivity. 
The conditions will be expressed by specifying the special positions the 
three wells must have inside a periodic cell.\\
It is worth mentioning that due to its topological aspect, it is natural to study the Chern index values in a generic situation, because topological properties are robust against perturbations. Secondly, in the generic ensemble of Hamiltonian under study, the integer values of the Chern indices are characterized by the position of the boundary where their value changes by one unit. These boundaries turn out to correspond to degeneracies in the spectrum \cite{chern3}. We are therefore brought to study the generic location of these degeneracies, in the space of Hamiltonians. This is done in section 5 and 6.\\
Under reasonable assumptions on the tunnelling interaction in phase space, we find that the boundary of constant Chern index domain form ellipses 
in any generic two-dimensional subspace of the Hamiltonian's space. 
These analytical results are well confirmed by numerical calculations 
in section 7.\\
These results extend previous work by the first author 
for the Hall conductivity resulting from the tunneling effect between two wells 
in a given periodic cell (\cite{fred1}, \cite{fredtsg}).
Although the methods looks similar,
calculations and results are quite different.

\section{Quantum mechanics on the torus}

\subsection{Classical dynamics on the torus}

Let us consider a one-degree of freedom Hamiltonian (hence an integrable
dynamics), periodic both in
position $q$ and momentum $p$, with respective periods $Q$ and $P:$%
\begin{equation}
H(q,p)=H(q+Q,p)=H(q,p+P)  \label{e:periodicity}
\end{equation}

The function $H(q,p)$ can be decomposed into its Fourier series: 
\begin{equation}
H(q,p)=\sum_{n_1,n_2\in Z^2}c_{n_1,n_2}\exp (i2\pi n_1\frac qQ)\exp (i2\pi
n_2\frac pP)  \label{e:hc}
\end{equation}
This decomposition will be used for numerical computations, but is not
essential for the theoritical part of this work.

Since $H$ is a real valued function, the complex coefficients $c_{n_1 n_2}$
must satisfy: 
\[
c_{n_1,n_2}=\bar c_{-n_1,-n_2}\in I\!\!\!\!C ,
\quad (n_1,n_2)\in Z\!\!\!Z ^2 
\]

The trajectories $q(t),p(t)$ evolve on the plane, but because of
periodicity, they can be considered as trajectories on the Torus $T_{qp},$
obtained by identifying the opposite sides of the cell $[0,Q]\times [0,P].$

\begin{figure}[htbp]
{\centering \epsfig{file=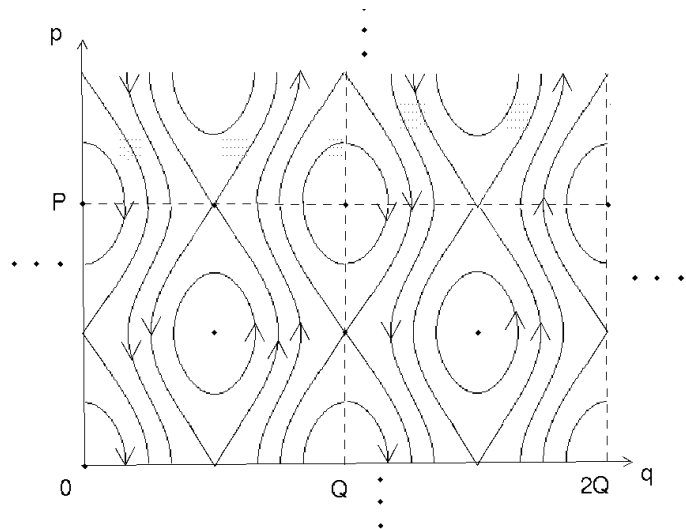, width=0.5\textwidth} \par}
\caption{Trajectories of the Harper model (\ref{eq:traj}). }
\label{fig:traj}
\end{figure}

For example, the trajectories of the Harper Hamiltonian 
\begin{equation}
\label{eq:traj}
H(q,p)=-\cos (2\pi \frac qQ)-\frac 12\cos (2\pi \frac pP) 
\end{equation}
are displayed in figure \ref{fig:traj}, p. \pageref{fig:traj}. 
Some energy levels are contractibles, others are not.

% fig1:traj

\subsection{Quantum mechanics on the torus}

The Hilbert space of a particle with one-degree of freedom is $L^2(I\!\!R )$,
with the fundamental operators of position $\hat q$ and momentum $\hat p$
acting on a wave function $\psi (x),$ also noted $|\psi >.$

We now choose a symmetric quantization procedure: to the classical
Hamiltonian $H$, we associate the quantum Hamiltonian:
\[
\hat H=\sum_{n_1,n_2\in Z^2}\frac 12c_{n_1,n_2}\exp (i2\pi n_1\frac{\hat q}%
Q)\exp (i2\pi n_2\frac{\hat p}P)+\mbox{hermitian conjugate} 
\]

We denote by $\hat{T} _Q$ [respectively $\hat{T}_P$] the translation
operator by one period. $\hat T_Q$ translates by $Q$ a wave function $\psi (x)$
and $\hat T_P$ translates its Fourier transform $\hat{\psi }(p)$ by $P$:
\begin{eqnarray}
 \hat{T}_Q \psi(x)&= &\psi(x-Q)   \label{eq:defTQ} \\ 
 \hat{T}_P \hat{ \psi}(p)&= &\hat{ \psi}(p-P)  \label{eq:defTP}
\end{eqnarray}
We may rewrite equations (\ref{eq:defTQ}) and (\ref{eq:defTP}) as:
\[
\hat T_Q=\exp (-iQ\hat p/\hbar ),\quad \hat T_P=\exp (iP\hat q/\hbar ) 
\]

Quantum mechanically speaking, the periodicity Eq. 
(\ref{e:periodicity}) reads: 
\begin{equation}
\lbrack \hat H,\hat T_Q]=[\hat H,\hat T_P]=0  \label{e:periodicity2}
\end{equation}

To continue, we now have to assume that 
\begin{equation} \label{eq:commutativite}
[\hat T_Q,\hat T_P]=0.
\end{equation}
It is easy to prove that:
\[ \hat{T}_Q \hat{T}_P=  e^{-i\frac{QP}{\hbar } }
\hat{T}_P \hat{T}_Q  \]
hence (\ref{eq:commutativite}) is fulfilled
if and only if there exists an integer $N$ such that:
\begin{equation} \label{eq:main_hyp}
N=\frac{QP}{2\pi \hbar }\in I\!\!N^{*} 
\end{equation}
This hypothesis (\ref{eq:main_hyp}) can be regarded as a geometric quantization 
condition, which states that
there is an integer number of Planck cells in the phase space. The 
semi-classical limit $\hbar \rightarrow 0$ corresponds to the limit
$N\rightarrow +\infty$.

\begin{rem} \label{rem:quantization}
For a given periodic Hamiltonian $H$ and a given Planck constant
$\hbar $, the ratio $\frac{PQ}{2\pi \hbar } $
is generically not an integer. But the spectral properties of $\hat{H}$
are in some sense continuous  with respect to $\hbar $, hence we
may approximate $\hbar $ by $\hat{\hbar } $
so that $\frac{PQ}{2\pi \hat{ \hbar} } $ becomes a rational $\frac{N}{D} $
(\cite{harper1}).
Now to fulfill our hypothesis, we consider $H$ as a periodic hamiltonian
with periods $(Q,D\times P)$ (or equivalently $(D\times Q,P)$, many other
combinations are possible if $D$ is not a prime number). In the
sequel, $P$ and $Q$ denotes periods of $H$ and not necessarily 
{\em primitive}\/ periods of $H$. From this, we see that it is natural
to expect additional translation symmetries of the Hamiltonian inside the 
(non primitive) periodicity cell and the main results of this work
will apply to the $D=3$ case.
\end{rem}

\begin{center}
We now assume that Hypothesis (\ref{eq:main_hyp}) is fulfilled.
\end{center}

According to the commutation relations (\ref{e:periodicity2}) and 
(\ref{eq:commutativite}),
the Hilbert space $L^2(I\!\!R )$ 
may be decomposed as a direct sum of the
eigenspaces of the translation operators $\hat T_Q$ and $\hat T_P$:
\begin{eqnarray}
L^2(I\!\!R ) &=&\int \int H_N(\theta _1,\theta _2)\ d\theta _1d\theta _2
\label{e:decomposition} \\
H_N(\theta _1,\theta _2) &=&\left\{ |\psi >
\quad \mbox{such that }\left\{ 
\begin{array}{c}
\hat T_Q|\psi >=\exp (i\theta _1)|\psi > \\ 
\hat T_P|\psi >=\exp (i\theta _2)|\psi >
\end{array}
\right. \right\}  \nonumber
\end{eqnarray}
with $(\theta _1,\theta _2)\in [0,2\pi [^2$ related to the periodicity of
the wave function under translations by an elementary cell. The space of the
parameters $(\theta _1,\theta _2)$ has also the topology of a torus, and
will be denoted by $T_\theta .$

The space $H_N(\theta _1,\theta _2)$ is not a subspace of $L^2(I\!\!R )$,
the space of physical states, it is a space of distributions
in the $x$ representation, we will see later that it is a subspace
of a weighted $L^2$
space in the Bargmann representation. We will now show that 
$H_N(\theta _1,\theta _2)$ is finite dimensional.
Let $|\psi >\in H_N(\theta _1,\theta_2)$.
The Fourier transform of $\psi (x)$ is $\theta _2$-Floquet-periodic
of period $P$, so $\psi(x)$ is discrete, it is a sum of Dirac 
distributions supported at points distant from $\frac hP=\frac QN$ 
from each other. 
Moreover, $\psi (x)$ is $\theta_1$-periodic, hence $\psi $ is characterized 
by the $N$ coefficients at the $N$ Dirac distributions supporting points 
in the interval $q\in [0,Q[$. 
Eventually we get:
\[
\dim _{I\!\!\!\!C} \ H_N(\theta _1,\theta _2)=N 
\]

Because of Eq. (\ref{e:periodicity2}) the Hamiltonian $\hat H$ is
block-diagonal with respect to the decomposition 
Eq. (\ref{e:decomposition}), 
so we have to consider the spectrum $\sigma (\theta _1,\theta _2)$
of $\hat H$. The operator $\hat{H}$ acts  
on $H_N(\theta _1,\theta _2)$ as a $N\times N$ hermitian matrix, its spectrum is made of $N$ eigenvalues:
\[ \sigma (\theta _1,\theta _2) =
\{ E_1(\theta _1,\theta _2), ..., E_N(\theta _1,\theta _2) \} \]
Let $\psi _1$, ..., $\psi _N$ be the corresponding eigenfunctions:
\[
\hat H|\psi _n(\theta _1,\theta _2)>=E_n(\theta _1,\theta _2)|\psi _n(\theta
_1,\theta _2)>\qquad n=1,..,N 
\]
For a given level $n,$ as $(\theta _1,\theta _2)\in T_\theta $ are varying,
the energy level $E_n(\theta _1,\theta _2)$ form a band, and (assuming that 
$E_n(\theta _1,\theta _2)$ is never degenerate $\forall \theta $), 
the eigenvectors 
$|\psi_n(\theta _1,\theta _2)>$ form a $2D$ surface in the quantum 
states space.
But for a fixed $(\theta _1,\theta _2),$ and any $\lambda \in I\!\!\!\!C $, 
$\lambda|\psi _n(\theta _1,\theta _2)>$ is also an eigenvector. 
So the family of
eigenvectors for the level $n$ form a complex-line-bundle (of
fibre $\cong I\!\!\!\!C \ni \lambda $ ) in the projective space of the bundle
$H_N \rightarrow T_\theta $. The
topology of this line bundle is characterized by an integer $C_n\in Z\!\!\!Z$,
called the Chern index. Because of the natural Hilbert scalar product on 
$L^2(I\!\!R )=\int \int H_N(\theta _1,\theta _2) \ d\theta _1 d\theta _2$, 
which induce the Berry (or Chern) connection (\cite{berry1}, \cite{chern6}), 
this topological
number is explicitly given by the integral of the Berry (or Chern)
curvature (\cite{iqhe2}):
\begin{equation}
C_n=\frac i{2\pi }\int \int
\left(< \partial _{\theta _1} \psi _n(\theta _1,\theta_2)
| \partial _{\theta _2}\psi _n(\theta _1,\theta _2)>
-< \partial _{\theta _2} \psi _n(\theta _1,\theta_2)
| \partial _{\theta _1}\psi _n(\theta _1,\theta _2)>\right)
\ d\theta _1 d\theta _2
  \label{e:chern}
\end{equation}
This expression has been used intensively for our numerical calculations.
Moreover, it can be shown (see e.g. \cite{iqhe1},\cite{chern4}) that:
\begin{equation} \label{eq:sigmacn}
\sum _{n=1}^N C_n=1
\end{equation}

\section{Husimi and Bargmann representation}
We have seen previously that the space $H_N$ is not a subspace of 
$L^2(I\!\!R )$. For functions belonging to $H_N$, it will be 
more useful to introduce a phase space representation
of the quantum states, called the Bargmann representation (\cite{bargmann}).

Consider a Quantum state $|\psi >\in L^2(I\!\!R )$. In order to
characterize the localization of $|\psi >$ in the phase space near 
the point $(q,p)$, we first construct a Gaussian wave packet $|qp>$
(coherent state) defined in the $x$-representation by:
\[
<x|qp>=\left( \frac 1{\pi \hbar }\right) ^{1/4}\exp (\frac
i\hbar px)\exp (-\frac{(x-q)^2}{2\hbar }) 
\]
The notation $|qp>$ recalls that the coherent state is localized (in the
semi-classical limit) at the point $(q,p)$ of the phase space.

The Husimi distribution of a state $|\psi >$ is defined over the phase space
by:
\[
h_\psi (q,p)=|<qp|\psi >|^2 
\]
and for $\varphi \in L^2(I\!\!R )$, we have:
\begin{equation} \label{eq:coherentL2}
 \int |\varphi (x) |^2 \ dx = \int \int |<qp|\varphi >|^2  \ \frac{ dq \ dp
}{2\pi \hbar }
\end{equation}

To characterize the functions of $L^2(T^{\star}I\!\!R )$ which are $(q,p)$ representations of a state, it is more
convenient to introduce a complex representation of the phase space
$z=\frac 1{\sqrt{2\hbar }}(q+ip)$.
Another (proportional) expression of the coherent state is then: 
\[ |z > = \exp (za^{+})|0> \]
with $|0>$ being the fundamental of the harmonic oscillator $H_0=\hat
q^2+\hat p^2,$ and $a^{+}$ being the associated creation operator.
Indeed:
\[ |qp > =  \exp (i\frac{qp}{2\hbar }-\frac{q^2+p^2}{4\hbar })|z> \]
The following antiholomorphic function of $z$ is called the Bargmann
distribution of $\psi $:
\[
b_\psi (z)=<z|\psi > 
\]

Clearly, we have 
\[ h_\psi (q,p)=\left| b_\psi (z)\right| ^2 e^{-\frac{q^2+p^2}{2h}} , \]
hence the zeroes of the function $h_\psi (q,p)$ are those of the 
holomorphic function $b_\psi (z)$, which are localized zeroes 
in the phase space. Moreover, (\ref{eq:coherentL2}) implies
that $\psi \in L^2(I\!\!R )$ if and only if
$b_\psi \in L^2(I\!\!\!\!C ,e^{-|z|^2/h})$ and $b_\psi $ is antiholomorphic.

The same definitions can be applied for a state $|\psi >\in H_N(\theta
_1,\theta _2)$. The corresponding Bargmann function is a theta-function 
\cite{zero2} and the Husimi distribution is bi-periodic in $(q,p)$, hence is
well defined on the
Torus $T_{qp}$.  In this representation,
$H_N$ is a subspace of $L^2_{loc}(I\!\!\!\!C ,e^{-|z|^2/h})$ (we keep the
weight $e^{-|z|^2/h}$ since it is the natural one in the Bargmann
representation).

\begin{rem} \label{rem:zerons}
The Bargmann and Husimi representation of a function of 
$H_N(\theta _1,\theta _2)$
are characterized by their $N$ zeroes in a given cell 
$T_{qp}$ up to a multiplicative constant. Since the $N$ zeroes are
constrained to have a fixed sum (\cite{zero2}), 
we get the right dimension $N-1+1$ for the Hilbert space 
$H_N(\theta _1,\theta _2)$.
\end{rem}

\section{Semi-classical expectation of the Chern index}
The question is now to guess the value of the Chern index of
a given band from classical informations. 
The first result in this direction is a characterization of the Chern index 
from the Husimi distribution \cite{chern2}:
\begin{proposition}
If there exists some point $(q,p)\in T_{qp}$ of the phase space, such that 
\[ \forall (\theta _1,\theta_2)\in T_\theta , \quad
h_{\psi _n(\theta _1,\theta _2)}(q,p) \neq 0 \] 
then $C_n=0$.
\end{proposition}
The proof is quite simple: if $<qp|\psi _n(\theta _1,\theta
_2)>$ is never zero then we can select an eigenstate $|\psi _n(\theta
_1,\theta _2)>$ in each fiber such that $\arg (<qp|\psi _n(\theta _1,\theta
_2)>)=0,$ giving us a non-vanishing section of the bundle. This section
is also a global frame, hence the bundle is trivial: $C_n=0$.

As a corollary, we get an important semi-classical result about Chern
indices:
\begin{proposition}
If the energy level $\Sigma _E=\{(q,p)$ such that $H(q,p)=E\}$ consists of
a single contractible trajectory, then the Chern index of the bands of
energy around $E$ are semiclassically zero (more precisely, if
we consider a $\hbar $-parametrized family of energy bands tending to
$E$ as $\hbar $ tends to 0, then for $\hbar $ sufficiently small, the
Chern index of the energy band must be 0).
\end{proposition}
Indeed the WKB construction of quasimodes in phase space 
(\cite{wkb1}, \cite{wkb2}, \cite{wkb3}) shows
that the Husimi distribution $h_{\psi _n(\theta _1,\theta _2)}(q,p)$ of the
eigenstates are highly concentrated and non-vanishing in the vicinity of the
trajectory. 
Thus, taking $(q,p)$ on the classical trajectory and from the proposition 2, we obtain $C_n=0$.

Hence, to get a non-zero Chern index, we must investigate situations
where the energy level $\Sigma _E$ is not connected or is non-contractible. 
In this
paper, we will study the first situation. The second one is slightly
different, but very interesting since it should explain the generic
non-zero Chern index arising from Eq. (\ref{eq:sigmacn}).

We need some results about
non-zero Chern indices. In fact, for any fixed point $(q,p)$, 
the Chern index $C_n$ is the algebraic number of
time that a zero of $h_{\psi _n(\theta _1,\theta _2)}(q,p)$ crosses
$(q,p)$ \cite{iqhe3}:
\begin{proposition}
Assume that for each $\theta \in T_\theta $, the eigenvalues 
$E_n(\theta _1,\theta _2)$ are non-degenerate. Let
$Z(\theta _1,\theta _2)$
be the set of the non-ordered $N$ zeroes of 
$h_{\psi _n(\theta _1,\theta _2)}$. Fix a point $(q,p)$ of the
phase space. Define:
\[ N(q,p)=\{ (\theta _1,\theta _2)\in T_\theta \ / \quad
(q,p)\in Z(\theta _1, \theta _2) \} \]
Then:
\begin{equation} \label{eq:chern}
C_n=\sum _{(\theta _1,\theta _2)\in N(p,q)} \pm 1 
\end{equation}
where the sign $\pm$ corresponds to the local orientation of the mapping $Z_i$
at $(\theta _1,\theta _2)$, where $Z=\{Z_1, ..., Z_N\}$ and 
$Z_i(\theta _1, \theta _2)=(q,p)$.
\end{proposition}
For example, if the energy level is made of two connected components 
$\Gamma _1$ and $\Gamma _2$, 
then from the tunnelling effect, the eigenstates $|\psi _n(\theta
_1,\theta _2)>$ are a superposition of quasimodes $|\psi _1>$ and $|\psi _2>$
localized on each trajectory. If this superposition is fluctuating enough
when $(\theta _1,\theta _2)$ are varying, then $h_{\psi _n(\theta _1,\theta
_2)}(q,p)$ can vanish for every point $(q,p)$ and we expect to get $C_n\neq 0$.
In fact this is possible only for special configurations of the two
trajectories. This has been investigated in detail in \cite{fred1}, showing
that in specific situations, we can observe $C_n=\pm 1$.
In the following sections, we will study the (more complicated) case of 
three contractible trajectories of energy $E$.

\section{Generic family of Hamiltonian $H_{(\gamma _1,...,\gamma _p)}$,
Chern indices and degeneracies.}
 As in \cite{fred1}, Eq. (\ref{eq:chern}) is not so useful to compute analytically the Chern indices 
of a given Hamiltonian $H$. The strategy we will adopt is to build a path of Hamiltonians 
$H(t), t\in [0,1]$ such that $H(0)=H$ and $H(1)$ has zero Chern indices (for $H(1)$ no tunnelling effect occurs so eigenfunctions are supported by only one connected and contractible trajectory in the phase space, hence Chern indices are zero). Along the path $H(t)$, Chern indices are constant because a continuous application with discrete values is constant. Exception is when a degeneracy occurs between eigenvalues. In this case the Chern index changes by $\pm 1$. In order to calculate the Chern indices,
we are therefore left to compute these degeneracies.

In this section, we will study a generic parametrized Hamiltonian family
(in other words, we will consider a submanifold of the space of Hamiltonians),
our main interest is to detect eigenvalues degeneracies.

There is an essential property for our investigations (\cite{chern5}):
\begin{proposition}
For a generic family of Hermitian matrices, degeneracies between two levels
of the spectrum occur with codimension 3.
\end{proposition}
This means for example that for a generic 3 dimensional 
family of Hermitian matrices 
$M(\gamma _1,\gamma _2,\gamma _3)$ the value of $\gamma _1,\gamma _2,\gamma_3$
for which two levels are degenerate, are points in the space of 
$(\gamma _1,\gamma _2,\gamma _3)$.

Consider now a parametrized family of classical Hamiltonian $H_\gamma$ on the torus with 
external parameters $\gamma=(\gamma_1,...,\gamma _p)$. 
The Fourier components $c_{n1,n2}$ of $\hat H$ in Eq. (\ref{e:hc}), 
the shape and position of each trajectory in the phase space depend
on these classical parameters $\gamma_1,...,\gamma _p$. 
On the quantum side, the matrix of the Hamiltonian in a specific base,
depends on the classical parameters $\gamma_1,...,\gamma _p$ and on the 
2 quantum parameters $\theta _1,\theta _2$.
Since degeneracies are of codimension three in the space 
$(\theta _1,\theta _2,\gamma_1,...,\gamma _p) $, they are of codimension $1$ (hypersurfaces) if we
project them onto the space of classical parameters $(\gamma _1,...,\gamma _p)$.
In fact for each point $(\gamma_1,...,\gamma _p)$ not belonging to such 
a hypersurface, the Chern index of a
given band $C_n$ may be calculated with Eq. (\ref{e:chern}).
If we cross a degeneracy hypersurface corresponding to the band
$n$ and $n+1$, the value of $C_n$ and $C_{n+1}$ changes respectively
by $\pm 1$ and $\mp 1$ (because the sum is conserved: cf. \cite{chern3}).
For example, in a one-dimensional space $\gamma $, degeneracies appear
as points. For a two-dimensional space, we summarize the previous results as:
\begin{proposition} 
In a two dimensional space $(\gamma _1,\gamma _2)$, degeneracies appear 
as lines bordering different values of the Chern index. The Chern index
changes by $\pm 1$ when crossing a line.
\end{proposition}
In the previous section,
we mentioned that these lines occur only if the tunnelling effect occurs. 
In the next section, we will determine (in the semi-classical
limit) the
location of these lines when the tunnelling effect occurs between three
trajectories in a periodic cell.

\section{Degeneracies due to the tunnelling effect between three trajectories of
energy $E$}

In this section, we consider a generic family of periodic
Hamiltonians $H_{\gamma}$. We
will use local coordinates $\gamma =(\gamma_1,...,\gamma _p)$.

\begin{figure}[htbp]
 \epsfig{file=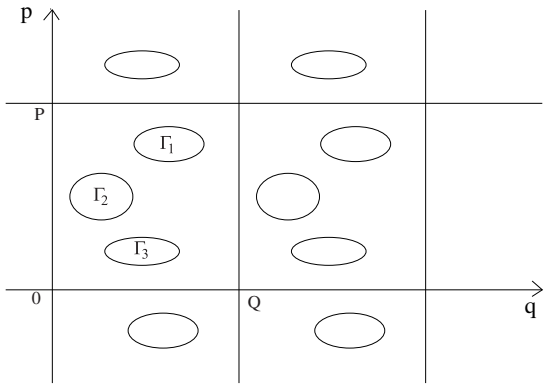, width=0.4\textwidth}
\caption{The energy level $E$ in the phase space.} 
\label{fig:3puits}
\end{figure}

We assume that for a given value of $E$, the intersection
of the periodic torus with the energy level $\Sigma _E$
consists of three contractible trajectories $\Gamma _i$
(figure \ref{fig:3puits}, p. \pageref{fig:3puits}):
\[ \Sigma _E \cap T_{qp}=
\{(q,p)\in T_{qp}/ H(q,p)=E\}=\Gamma _1\cup \Gamma _2\cup \Gamma _3 \]
We do not assume that $E$ is non-critical since construction of
quasimodes along critical trajectories may be done as in \cite{CPI}, \cite{CPII}. 
The only characteristic  we will use is that the mean distance between energy levels is of order $\hbar / |\ln \hbar| $ for a critical trajectory, whereas it is of order $\hbar$ for a non-critical trajectory.

% fig2:3puits

Our purpose is now to
establish under which conditions (shape and locations of the trajectories $\Gamma _i$)
degeneracies in the spectrum of $\hat H$ may occur semi-classically
near the energy $E$.
More precisely, we will describe the generic degeneracy lines in the
space $\gamma $.

\subsection{The tunnelling interaction matrix.} \label{sec:interaction}

In this section, we construct a basis of $H_N(\theta _1,\theta _2)$
and describe the asymptotic of the matrix of the Hamiltonian $H$ in this
basis. To have a more precise picture of these asymptotics, we will
make assumptions on the respective decays of interaction terms. 

For a first reading, one can skip this section and refer to the results mentionned in proposition \ref{prop:interaction}.

%To make reasonable assumptions, we remind the reader of the decay properties
%of the eigenvectors of Schr\"odinger operators 
%(in other words we will make assumptions on
%microlocal tunnelling effect which are fulfilled for 
%the tunnelling effect: see e.g.
%\cite{martinez94}, \cite{nakamura95} for partial results in this direction).

Using the ellipticity of $\hat{H} -E$ outside the classical region
$H(q,p)=E$, it is easy to prove that the eigenfunctions are of order
$O(h^N)$ for all $N\in I\!\!N $ (noted $O(h^{\infty})$) outside the classical region.
%(In the Schr\"odinger situation, the decay is more precisely exponential.)
This property is in fact sufficient to construct a basis of 
$H_N(\theta _1,\theta _2)$. 

\begin{figure}[htbp]
 \epsfig{file=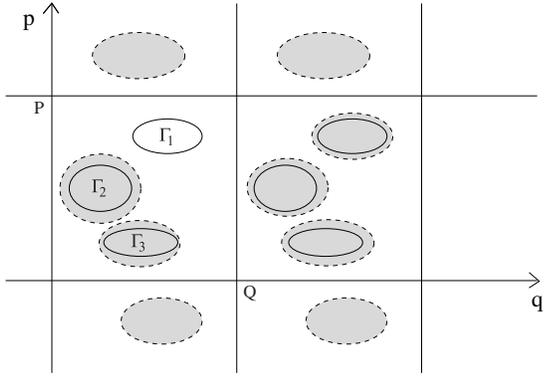, width=0.4\textwidth}
\caption{Modification of $H$.} 
\label{fig:mod3puits}
\end{figure}

It is easy to modify the Hamiltonian $H$ in a new hamiltonian
$\tilde{H}_1$ such that:
\begin{itemize}
\item the energy shell $E$ of $\tilde{H}_1$ is $\Gamma _1$,
\item $H=\tilde{H}_1$ outside the shaded region, see 
figure \ref{fig:mod3puits}, p.\pageref{fig:mod3puits}.
\end{itemize} 

%fig3:mod3puits

Let $|1>$ be an eigenfunction of $\tilde{H}_1$ corresponding to
an eigenvalue $\tilde{E}_1(h)$ such that
$\tilde{E}_1(h)\rightarrow E$ as $h\rightarrow 0$. 
Then $(\hat H -\tilde{E}_1 )|1>= (\hat H - \tilde{H}_1 )|1>$. In the last expression, the operator is  microlocally supported in the shaded region where $|1>$ is not. Hence:
\[ (\hat H -\tilde{E}_1 )|1> = O(h^\infty ) \]
%In the Schr\"odinger case, we would get under generic assumptions:
%\[ (\hat H -\tilde{E}_1 )|1> = O(h^{-N}e^{-S/h} ) \]
%for some integer $N$, where $S$ is the Agmon distance between $\Gamma _1$
%and $\Gamma _2 \cup \Gamma _3$.
The same construction applies for the trajectories $\Gamma _2$ and $\Gamma _3$
and we get the functions $|2>$ and $|3>$ microlocally supported
on $\Gamma _2$ and $\Gamma _3$ respectively .
In the sequel, we will denote by $r$ the greatest of the quantities
$| (\hat H -\tilde{E}_i )|i>|$ ($i=1,2,3$).
 %The reader may replace
%$r$ by $O(e^{-S/h})$ which is conjectured to be the right decay order,
%but in the following calculations it is sufficient to know that 
%$r=O(h^\infty )$.

\begin{figure}[htbp]
 \epsfig{file=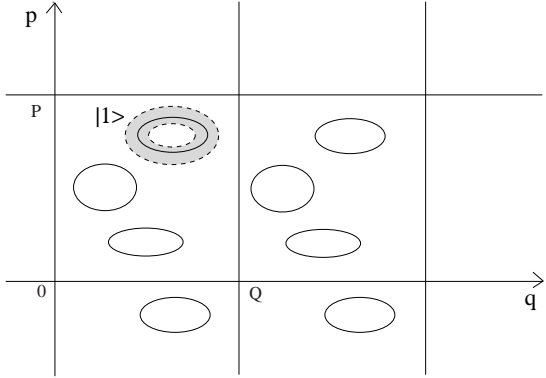, width=0.4\textwidth} 
\caption{Microlocal support of the states 
$|1>$ in the phase space.} 
\label{fig:state1}
\end{figure}

\begin{figure}[htbp]
 \epsfig{file=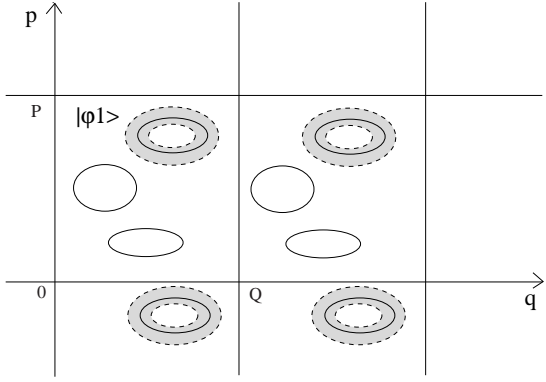, width=0.4\textwidth} 
\caption{Microlocal support of the state $|\varphi _1 >$ in the phase space.} 
\label{fig:statephi1}
\end{figure}

>From $|1>$, $|2>$ and $|3>$ we construct Floquet-periodic quasimodes:
\begin{equation}
|\varphi _i> =P(\theta _1,\theta _2)|i>\mbox{ for }i=1,2,3  \label{e:periode}
\end{equation}
where $P(\theta _1,\theta _2)$ is the operator from $L^2(R)$ to $H_N(\theta
_1,\theta _2)$ which makes a given state Floquet-periodic: 
\begin{equation} \label{eq:Ptheta}
P(\theta _1,\theta _2) := \sum_{n_1,n_2\in Z^2}\exp (-in_1\theta
_1-in_2\theta _2)T_Q^{n_1}T_P^{n_2} 
\end{equation}
Figure \ref{fig:state1} p.\pageref{fig:state1} and \ref{fig:statephi1} p.\pageref{fig:statephi1}
shows the microlocal support of $|1>$ and $|\varphi _1>$.

%fig4:state1
%fig5:statephi1

%Equation (\ref{eq:Ptheta}) is the Fourier serie
%of $P(\theta _1,\theta _2)\varphi $ (which is periodic of period $2\pi $
%with respect to $\theta _1$ and $\theta _2$), hence the first coefficient
%of the Fourier serie is:
%\[ \varphi =\frac{1}{(2\pi )^2} \int _{T_\theta } 
%P(\theta _1,\theta _2)\varphi  \ d\theta \]
%and by Parseval's Theorem:
%\begin{equation} \label{eq:parseval}
%\frac{1}{(2\pi )^2} \int _{T_\theta }
%|P(\theta _1,\theta_2)\varphi|^2 \ d\theta =
%\sum_{n_1,n_2} |T_Q^{n_1}T_P^{n_2} \varphi |^2
%\end{equation}
%In this representation (\ref{eq:parseval}) implies:
%\[ \frac{1}{(2\pi )^2} \int _{T_\theta }
%|P(\theta _1,\theta_2)\varphi|_{H_N}^2 \ d\theta = 
%\int _{z \in [0,Q]+i[0,P]} \sum_{n_1,n_2} |T_Q^{n_1}T_P^{n_2} \varphi |^2(z) 
%e^{-|z|^2} \ dz \ d\overline{z}
%= ||\varphi ||^2
%\]

Suppose that we are in the resonance situation: $\tilde{E}_i(\hbar )-\tilde{E}_j(\hbar )=o(\hbar )$ for $i,j=1,2,3$. 
%(if the three trajectories are symmetric, this assumption is automatically
%fulfilled since we may construct the modified hamiltonians
%$\hat{\tilde{H}_i}$ by translation,
%hence $\tilde{E}_i=\tilde{E}_j$). 
Then for $\varepsilon >0 $ sufficiently small the spectrum of $\hat{H} $ in the interval $I=[\tilde{E}_1(\hbar )-\varepsilon \hbar ,
\tilde{E}_1(\hbar )+\varepsilon \hbar ]$ is made of 3 eigenvalues
$E_1(\hbar )$, $E_2(\hbar )$ and $E_3(\hbar )$, and
if we denote by $\Pi _I(\theta )$ the corresponding spectral projector of 
$H_N(\theta )$, the spectral space is spanned by:
\[ |\psi _i>= \Pi _I(\theta ) |\varphi  _i > , \quad i=1,2,3.\]
Note that for a critical trajectory the width of the interval must be $\varepsilon \hbar / |\ln  \hbar |$.

Let $C$ be the circle of center $\tilde{E}_1(\hbar )$ and radius 
$\varepsilon \hbar $ in the complex plane. We represent $\Pi _I$ and identity
as:
\[ \Pi _I= \frac{1}{2\pi i} \int_C (\hat{H} -z)^{-1} \ dz,
\quad  Id= \frac{1}{2\pi i} \int_C (\tilde{E}_1-z)^{-1} \ dz \quad Id, \]
and get:
\begin{eqnarray*} 
|\psi _i>-|\varphi  _i> &= & \frac{1}{2\pi i} \int_C 
[ (\hat{H} -z)^{-1}-(\tilde{E}_1-z)^{-1} ] |\varphi _i>\ dz \\
&=& \frac{1}{2\pi i} \int_C
- (\hat{H} -z)^{-1}(\tilde{E}_1-z)^{-1} (\hat{H} -\tilde{E}_1) 
|\varphi _i>\ dz \\
&=& O(r/h^2) (=O(h^\infty ) )
\end{eqnarray*}
since $||(\hat{H} -z)^{-1}||<K/(\varepsilon \hbar )$
and $ |\tilde{E}_1-z|^{-1}<K/(\varepsilon \hbar )$ for 
$z\in C $.
In the following text, we replace $r/h^2$ by $r$, the new $r$ function has
decay properties similar to the original one ($r=O(h^\infty )$ and is
probably exponentially decaying).

Hence $|\psi _i>$ is microlocalized on $\Gamma _i$ like 
$|\varphi _i>$.

The next step is to get an orthonormalized basis from the $(|\psi _i>)_i$.
Let
\begin{equation} \label{eq:N}
n_{ij}=<\psi _j|\psi _i> , \quad N=(n_{ij})_{i,j}
\end{equation}
Since $N$ is positive hermitian, there exists
a unique positive hermitian matrix $B$ such that:
\begin{equation} \label{eq:defB}
B^{-2}=N
\end{equation}
Let
$B=(\beta  _{ij})_{i,j}$ (with $\beta  _{ij}=\overline{\beta  _{ji}}$)
and:
\begin{equation} \label{eq:B}
|e_i>=\sum _j  \beta  _{ij} |\psi _j>
\end{equation}
We claim that the family  $|e_i>$ is orthonormal. Indeed:
\begin{eqnarray*} 
<e_i|e_j>&=&\sum _{k,l} \overline{\beta  _{ik}} \beta  _{jl} <\psi _k|\psi _l>
\\
&=& \sum \beta  _{jl} n_{lk} \beta  _{ki}\\
&=& ( B N B )_{ji}\\
&=& (Id) _{ji} 
\end{eqnarray*}

The tunnelling interacting matrix is the $3\times 3$ Hermitian matrix: 
\begin{equation}
A=(<e_i|\hat{H}|e_j>)_{ij} 
\end{equation}
The eigenvalues of $A$ are exactly the eigenvalues of $\hat{H}$ which belong to the $I$ interval.

We will now compute the semi-classical asymptotics of $a_{ij}$.
Let:
\begin{equation} 
G=(<\psi _i| \hat H |\psi _j>)_{i,j=1,2,3}  \label{e:A}
\end{equation}
>From Eq. (\ref{eq:B}), we get:
\begin{equation} \label{eq:calculA}
A=\overline{B} G \overline{B}
\end{equation}
hence we want to compute asymptotics of $B$ and $G$.

First we remark that in Eq.(\ref{eq:N}), 
we can replace $|\psi >$ by $|\varphi >$, since by Pythagoras theorem:
\begin{equation} \label{eq:pythagore}
<\varphi _i|\varphi _j>= <\psi _i|\psi _j>
+<\varphi _i-\psi _i|\varphi _j-\psi _j>
= <\psi _i|\psi _j> + O(r^2 ) 
\end{equation}
Since $\hat{H}$ commutes with $\Pi _I$, we get by the same method:
\[ <\varphi _i|\hat{H}|\varphi _j>= <\psi _i|\hat{H}|\psi _j>
+<\varphi _i-\psi _i|\hat{H}|\varphi _j-\psi _j>
= <\psi _i|\hat{H}|\psi _j> + O(r^2 ) \]
 We have
\[ <\varphi _i|\varphi _j>=\delta _{ij} + O(r) \]
Hence: 
\begin{equation} \label{eq:asymptNB}
N=Id+O(r), \quad B=Id+O(r)
\end{equation}

Now, we want to expand the functions $|\varphi _i>$ using (\ref{e:periode})
and (\ref{eq:Ptheta}) and evaluate the scalar products.
% This can be done
%in the Bargmann representation. Indeed functions of $H_N$ are characterized
%by their value on $[0,Q]\times [0,P]$ and
%the scalar product is inherited from the standard 
%$L^2([0,Q]\times [0,P],e^{-|z|^2/\hbar })$ 
%scalar product.
>From Eq.(\ref{e:A}) and (\ref{e:periode}), we get: 
\begin{equation} \label{e:a11} 
<\psi _1|\hat H|\psi _1> = <1|\hat H|1>+2\sum_{\vec{n}\in D}Re\left( \exp (-i \vec{n}\vec{\theta} )<1|\hat H|1_{\vec{n}}>\right)
\nonumber
\end{equation}
where:
\begin{itemize}
\item $\vec{n}=(n_1,n_2)$,  $\vec{\theta} =(\theta _1,\theta_2)$,
\item $D=\{(Z\!\!\!Z \times I\!\!N )\setminus (-I\!\!N \times \{0\})\}$ 
is the half plan of $\vec{n}$,
\item 
$|1_{\vec{n}}>=T_Q^{n_1}T_P^{n_2}|1>$ is a quasi mode concentrated on the 
translated trajectory $(\Gamma _1)_{n_1,n_2}$ in the $(n_1,n_2)$ cell 
of the phase space.
\end{itemize}

The first term of (\ref{e:a11}) is $\tilde{E}_1+O(r^2)$.
The term $<1|\hat H|1_{\vec{n}}>$ is of order $O(r)$ 
because it corresponds to the
tunnelling interaction between the quasimode $|1>$ localized in cell $(0,0)$
and the quasi mode $|1_{\vec{n}}>$ 
localized in the cell $(n_1,n_2)\neq (0,0)$ (we may need to modify the function
$r$ to get this decay: $r$ denotes the strongest tunneling
interaction between two components of the energy shell). 
Hence:
\[
<\psi _1|\hat H|\psi_1>=\tilde{E}_1 + O(r) 
\]
Similar asymptotics hold for the other diagonal terms.

A non-diagonal term of $G$ is e.g.: 
\begin{equation}
<\psi _1|\hat H|\psi _2> = \sum_{\vec{n}\in Z\!\!\!Z ^2}\exp (-i \vec{n}\vec{\theta} )<1|\hat H|2_{\vec{n}}>
\end{equation}
And there is generically only one leading term due to the strongest
tunnelling interaction between $|1>$ and $|2_{\vec{n}_{12}}>$ (located in the cell 
$\vec{n}_{12}$: $\vec{n}_{12}$ may be for example $(0,0)$ or $(0,\pm 1)$ or $(\pm 1,0)$), 
of strength 
\[ <1|\hat H|2_{\vec{n}_{12}}>=m_{12}\exp (i\varphi _{12}) \]
where $m_{12}>0$ is $O(r)$.
This gives: 
\[
<\psi _1|\hat H|\psi _2> \equiv m_{12}\exp (i(\varphi _{12}-\vec{n}_{12}\vec{\theta} )) 
\]
and similar expressions for others non-diagonal terms. Hence all non-diagonal terms are $O(r)$. If we denote by $\tilde{r}$ the second strongest
tunneling interaction, we have
\[
<\psi _1|\hat H|\psi _2> = m_{12}\exp (i(\varphi _{12}-n_{12}\theta )) + O(\tilde{r}) 
\]
We assume of course that $\tilde{r}=O(h^\infty )r$.
% in fact one would
%conjecture that $\tilde{r}$ decays exponentially with respect to $r$.

>From $A=\overline{B} G\overline{B}$, 
$B =N^{-1/2}$ and $N=Id+O(r)$,
we get for the diagonal terms of $A$:
\[ a_{ii}=\tilde{E}_i+O(r) \]
and for the non-diagonal terms:
\begin{equation} \label{eq:non_diag}
 a_{ij} =  m_{ij}\exp (i(\varphi _{ij}-\vec{n}_{ij}\vec{\theta} )) + O(\tilde{r})
\end{equation}

To get a more accurate asymptotic of the diagonal terms, we
will now study the shifted interaction matrix:
\[ <e_i|\hat{H} -\tilde{E}_3|e_j>= \overline{B} (G - \tilde{E}_3 N )  
\overline{B}\] 
We have $G-\tilde{E}_3 N = (<\psi _i|\hat{H} -\tilde{E}_3|\psi _j>)_{ij}$.
For the non-diagonal terms, we get similar asymptotics as above
(\ref{eq:non_diag}) with new constants. For simplicity, we will however 
keep the same notations, since the decays remain of the same order and the
phases were not explicit.
For the diagonal terms, we get by the same method:
\begin{equation} \label{eq:diag_terms}
(<\psi _i|H-\tilde{E}_3|\psi _i>)=\tilde{E}_i-\tilde{E}_3
+ m_{ii} \cos(\varphi _{ii}-n_{ii}\theta ) + ...
\end{equation}
Since $B=I$ modulo $O(r)$ and $G-\tilde{E}_3 N$ is
of order $O(r)$, the shifted interaction matrix is $G-\tilde{E}_3N$
modulo an error of order $O(\tilde{r})$.

In this section we have used he fact that the tunnelling interaction gives terms of order 
$r=O(h^\infty)$. In fact the reader may replace
$O(h^\infty)$ by $O(e^{-S/h})$ which is conjectured to be the right decay order, where $S$ is similar to the Agmon distance between two trajectories (as in the Schr\"odinger case $H=p^2/2+V(x)$). See \cite{martinez94}, \cite{nakamura95} for partial results in this direction).

We summarize the results of this section:
\begin{proposition} \label{prop:interaction}
Let $H$ be a Hamiltonian and $E$ an energy such that the energy shell
$H=E$ is made of three contractible connected components.
Let $|1>$, $|2>$, $|3>$ be the quasimodes of $\hat{H}$ localized near one of the 
three classical trajectories corresponding to eigenvalues 
$\tilde{E}_1(\hbar )\rightarrow E$, 
$\tilde{E}_2(\hbar )\rightarrow E$, $\tilde{E}_3(\hbar )\rightarrow E$ 
%of the Hamiltonians $\tilde{H}_i$ modified as explained at the beginning
%of this section. 
Suppose that 
$\tilde{E}_i(\hbar )-\tilde{E}_j(\hbar )=o(\hbar )$.
Then for $\varepsilon >0$ and $\hbar >0$ sufficently small,
the spectrum of $\hat{H}$ acting on $H_N(\theta _1,\theta _2)$
in the interval $[\tilde{E}_3(\hbar )-\varepsilon \hbar ,
\tilde{E}_3(\hbar )+\varepsilon \hbar ]$ is made of three eigenvalues.

We have constructed an orthonormal basis $|e_i >$ spanning the corresponding
3-dimensionnal vector space. Let $r$ be a majorant of the largest
tunneling interaction between two different trajectories, and $\tilde{r}$ be a majorant of the 
second largest tunneling interaction
 between two wells ($r=O(\hbar ^\infty )$ and $\tilde{r}
=O(h^\infty r)$. More precisely,
we choose $i$ and $j$ in$[1,3]$, we look at the second strongest
interaction between trajectory $i$ and translated of the trajectories $j$,
excluding the 0 translation if $i=j$.
$r$ and $\tilde{r}$ are conjectured to be exponentially decreasing with
respect to $r$). 
Then the matrix of $\hat{H}$ in the basis $|e_i >$ has the following 
asymptotic:
\begin{itemize}
\item The non-diagonal terms are $O(r)$, more precisely:
\[ a_{ij}=m_{ij } \exp{i(\varphi _{ij}-\vec{n}_{ij}.\vec{\theta} )} + O(\tilde{r}),
\quad m_{ij}=O(r) \]
\item The diagonal terms are given by:
\[ a_{ii} = \tilde{E}_i + m_{ii} \cos( \varphi _{ii}-\vec{n}_{ii} .\vec{\theta} ) +
O(\tilde{r}) \]
\end{itemize}

%\end{proposition}

%\begin{rem} \label{rem:gammadependance}
If we look at the dependency of the interaction matrix with respect to the external parameters $\gamma$, it is easy
to prove that:
\[ d_\gamma a_{ij}= (d_\gamma \tilde{E}_i) \delta _{ij}  + O(r) \]
where $d_\gamma$ denote the differential. The first term is dominant, since the variation of $\tilde{E}_i$ is of order $\hbar $ (or $\hbar /| \ln\hbar  |$ for a critical trajectory).

Since we are interested in eigenvalue degeneracies, we may substract 
$\tilde{E}_1.I_3$ to the interaction matrix, hence the variation of
the shifted interaction matrix with respect to $\gamma $
is described by the variation of  $\tilde{E}_1-\tilde{E}_3$
and $\tilde{E}_2-\tilde{E}_3$ with respect to $\gamma $ up to an error
of order $r$.
Hence, we will reduce our parameter space to two parameters $\gamma _1$
and $\gamma _2$, and we justify in appendix \ref{sec:parameters} that they  can be chosen as:
\begin{eqnarray} \label{e:paramm}
\gamma_1 &=& (\tilde{E}_1-\tilde{E}_3)/ \hbar \\
\gamma_2 &=&  (\tilde{E}_2-\tilde{E}_3)/ \hbar \nonumber
\end{eqnarray}

(replace $\hbar $ by $\hbar /|\ln\hbar | $ for a critical trajectory).

Moreover, since the error is of order $O(r)$, the degeneracy lines
in the parameter space will be of $O(\hbar ^\infty )$ size (and conjectured
to be of exponentially small size).
% hence after scaling we can forget
%that we are working with a manifold of Hamiltonian and act as if the $\gamma $
%dependency of the $H$ family was linear. 
We will come back to this point more
precisely in section \ref{sec:degeneracies}.
%\end{rem}
\end{proposition}

%==================================================================
%fred nov 97
\subsection{Model of a 3$\times$3 interaction matrix, and computation of its Chern indices.} \label{sec:matrice33}

This paragraph is self-contained. From the previous paragraph, we have to consider a continuous mapping on the torus $ T_\theta = [0,2 \pi[^2$ into the  3$\times$3 Hermitian matrices:
 \[ {\cal A}:(\theta _1,\theta _2) \in T_\theta  \rightarrow A(\theta _1,\theta _2)=(a_{i,j})_{i,j =1,2,3}
\]
where the diagonal terms are:
\begin{equation} \label{eq:diag}
a_{ii}=\tilde{E_i}+m_{ii}\cos ( \varphi_{ii} -\vec{n}_{ii} \vec{\theta} )\hspace{1cm} i=1,2,3
\end{equation}
and the non-diagonal terms are
\begin{equation} \label{eq:ndiag}
a_{ij}=m_{ij} \exp i (\varphi_{ij} -\vec{n}_{ij} \vec{\theta} ) \hspace{1cm} i<j =1,2,3
\end{equation}
with $\vec{n}_{ij}=(n_{1(ij)},n_{2(ij)})\in Z^2$ and $\vec{\theta}=(\theta _1,\theta _2)$.
We consider  $(m_{ij},\varphi_{ij},\vec{n}_{ij})_{i,j=1,2,3}$ as fixed. We will study only the dependance of $\cal A$ with the parameters  $(\tilde{E_1},\tilde{E_2},\tilde{E_3})$.

\begin{figure}[htbp]
 \epsfig{file=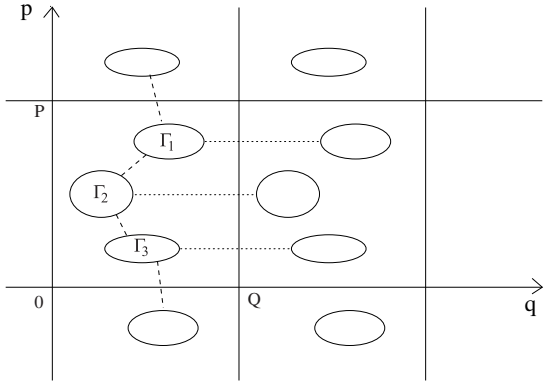,width=0.4\textwidth}
\caption{Example of tunnelling interaction. $\vec{n}_{12}=\vec{n}_{23}=0$, $\vec{n}_{31}=(0,1)$ and $\vec{n}_{11}=\vec{n}_{22}=\vec{n}_{33}=(1,0)$.}
\label{fig:interaction}
\end{figure}

We recall that  $(m_{ij},\varphi_{ij},\vec{n}_{ij})$  refer to  the tunnelling interaction between the three trajectories $(\Gamma _i)_i,i=1,2,3$  as sketched in figure (\ref{fig:interaction}), and  that $\tilde{E_i}$ is the energy of the quasimodes on trajectory  $\Gamma _i$.

If no degeneracy occurs in the spectrum of $A(\theta _1,\theta _2)$ (for all $\vec{\theta}=(\theta _1,\theta _2) \in T_\theta$), each eigenvector family has a well defined Chern index. Precisely, each eigenvector family $(|\psi_n(\vec{\theta})>)_{\vec{\theta}}$ for $n=1,2,3$ is a submanifold of the projective space $P(I\!\!\!\!C^3)$, homeomorphic to $T_\theta$. The complex line bundle structure of  $P(I\!\!\!\!C^3)$ induces a complex  line bundle over this submanifold whose topology is characterized by its Chern index.
For specific values of the parameters (codimension 1 in the space of $(\tilde{E_i})_i$) degeneracies occur and this causes a change of the Chern index. In this paragraph we will compute these Chern indices and the locus of the degeneracies in the space of $(\tilde{E_i})_i$ for each family of eigenvector $(|\psi_n(\vec{\theta})>)_{\vec{\theta}}$ for $n=1,2,3$.
First we remark that substracting the diagonal matrix $\tilde{E_3}Id$ to $A$ does not change its eigenvectors, so the only non-trivial external parameters are:
\begin{eqnarray} \label{eq:gammaE}
\gamma_1 &=& (\tilde{E}_1-\tilde{E}_3)/ \hbar  \\
\gamma_2 &=& (\tilde{E}_2-\tilde{E}_3)/   \hbar \nonumber
\end{eqnarray}
The locus of the degeneracies we are looking for are lines in the space $(\gamma_1, \gamma_2)$.

A general property of $3\times 3$ Hermitian matrices, mentioned in the
appendix \ref{sec:degen3}
is that a degeneracy occurs for the matrix $A$ if and only if: 
\begin{eqnarray}
\Im(a_{12}a_{23}a_{31}) &=&0  \label{e:degen} \\
a_{11}-a_{22}+\frac{a_{12}a_{23}}{a_{13}}-\frac{a_{21}a_{13}}{a_{23}} &=&0 
\nonumber \\
a_{22}-a_{33}+\frac{a_{13}a_{32}}{a_{12}}-\frac{a_{12}a_{23}}{a_{13}} &=&0 
\nonumber
\end{eqnarray}
$\Im$ stands for imaginary part.
Moreover, the degeneracy is between levels $n=1,2$ (respect. $n=2,3$) if $\Re(a_{12}a_{23}a_{32})>0$ (respect $<0$). 

The first equation  can be written: 
\begin{equation}
\Theta = \arg (a_{12}a_{23}a_{31})=\varphi _{12}+\varphi _{23}+\varphi
_{31}-(\vec{n}_{12}+\vec{n}_{23}+\vec{n}_{31})\vec{\theta} \equiv 0\qquad [\pi ]  \label{e:a}
\end{equation}

The phase $\Theta $ can be seen as the total tunnelling phase for the cycle of
trajectories $(\Gamma _1, \Gamma _2, \Gamma _3)$. We want to find solutions in $\vec{\theta}$ of this equation for fixed values of   $\varphi_{ij},\vec{n}_{ij}$. For that purpose, { \em we assume that } $(\vec{n}_{12}+\vec{n}_{23}+\vec{n}_{31})\neq 0$. { \em This is a generic assumption}. This means that
the cycle of trajectories $(\Gamma _1, \Gamma _2, \Gamma _3)$  is not contractible on the torus 
$T_{qp}$.

To simplify notations, we will assume that $\vec{n}_{12}=\vec{n}_{23}=0$, $\vec{n}_{31}=(0,1)$ and $\vec{n}_{ii}=(1,0)$. This corresponds to the tunnelling interactions sketched with dashed lines in figure (\ref{fig:interaction}).  For the tunnelling problem  considered in this paper, we think (without proof) that this case is the general one after a suitable lattice transformation in $SL(2,Z)$. But for the self-contained problem of this paragraph, the general case is solved in the
appendix \ref{sec:lissajou}.

%fig6:interaction

Eq. (\ref{e:a}) has then two solutions $\theta _2\equiv
\varphi _{12}+\varphi _{23}+\varphi _{31}\qquad [\pi ].$

Let
\[ s=\exp (i \Theta )=\pm 1 .\]

The two-last equations of Eq. (\ref{e:degen}) give:

\begin{equation} \label{eq:newdeg}
\tilde{E}_1 - \tilde{E}_3  = 
s m_{13} \left( \frac{ m_{12} }{ m_{23} } - \frac{ m_{23} }{ m_{12} }  \right)
+ m_{33} \cos(\varphi _{33}-\theta_1 )
- m_{11} \cos(\varphi _{11}-\theta_1 )
\end{equation}
and similarly for $\tilde{E}_2-\tilde{E}_3$.

Let us remark that:
\begin{eqnarray*}
A \cos(\theta_1 -\alpha )-B\cos(\theta_1 -\beta ) &=&
(A\cos \alpha - B\cos \beta ) \cos \theta_1 +
(A \sin \alpha - B\sin \beta )\sin \theta_1 
  \\
&=& D \cos(\theta_1 -\eta)
\end{eqnarray*}
where:
\begin{eqnarray} \label{eq:eta}
 D&=&\sqrt{A^2+B^2-2AB\cos(\alpha -\beta )}, \\
(\cos \eta, \sin \eta )&=&\left(
\frac{ A\cos \alpha - B\cos \beta }{D} ,
\frac{ A \sin \alpha - B\sin \beta }{D}
\right)  \nonumber
\end{eqnarray}

\begin{figure}[htbp]
 \epsfig{file=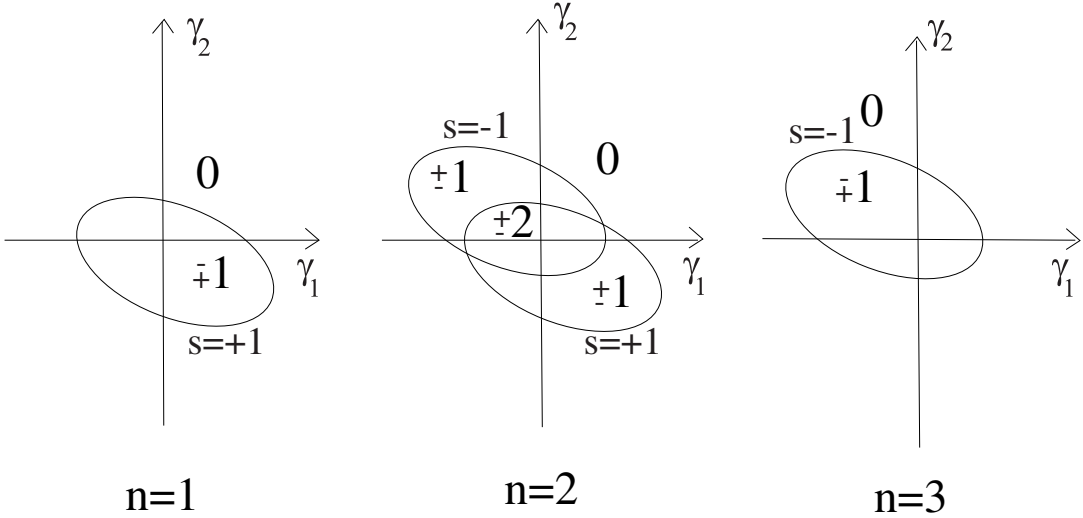, width=0.8\textwidth} 
\caption{Degeneracies lines and value of Chern indices from Eq.(\ref{eq:ellipses}).}
\label{fig:ellipses}
\end{figure}

So we put:
\begin{eqnarray} \label{eq:scales}
K_1 &=& \sqrt{ m_{11}^2+m_{33}^2-2m_{11}m_{33}
\cos \varphi _1 } 
\\
K_2 &=& \sqrt{  m_{22}^2+m_{33}^2-2m_{22}m_{33}
\cos \varphi _2 } \nonumber
\end{eqnarray}
where:
\[ \varphi _i=\varphi _{ii} - \varphi _{33} \]
and define:

\[
Y_1 = m_{13}
    \left( \frac{ m_{12} }{ m_{23} } - \frac{ m_{23} }{ m_{12} }  \right) \quad 
Y_2 = m_{23}
 \left( \frac{ m_{12} }{ m_{13} } - \frac{ m_{13} }{ m_{12} }  \right) 
\]
{ \em From (\ref{eq:newdeg}) and (\ref{eq:gammaE}), 
the degeneracy lines in the space $(\gamma_1,\gamma_2)$ are the 
following $t\in I\!\!R$ parametrized curves ($s= \pm 1$):
\begin{eqnarray} \label{eq:ellipses}
 \gamma_1(t)&=&(K_1 \cos(t-\eta) -s Y_1)/ \hbar\\
  \gamma_2(t)&=&  (K_2 \cos(t) -s Y_2) / \hbar \nonumber
\end{eqnarray}
where $s=\pm1$, $t=\theta_1-\eta_2$, $\eta=\eta_3 - \eta_2$ and the angles $\eta_2$ and $\eta_3$ may be computed using (\ref{eq:eta}).
}\\
As $t$ describes $[0,2\pi ]$, $(\gamma_1(t),\gamma_2(t))$ describes two translated ellipses of axes parallel to $\gamma_1=\pm \gamma_2$, one for each $s\in \{-1, 1\}$. $s=1$ (respect. $s=-1$) gives the degeneracy line between levels $n=1,2$ (respect. levels $n=2,3$). See figure (\ref{fig:ellipses}). These two ellipses may intersect, but they don't really intersect in the  
whole space $(\theta _1,\theta _2,\gamma_1,\gamma_2)$ because they correspond to two
different values of $\theta_2$.

Outside the ellipses, the Chern indices are zero. This is because, when $\gamma_1,\gamma_2 \rightarrow \infty$ the $A$ matrix goes to a diagonal matrix with trivial eigenvectors. Crossing a degeneracy line changes a Chern index change by $\pm 1$. It can be shown that the value $\pm 1$ is related to the orientation relatively to the orientation of the parametrized ellipse. From this, we deduce the values of the Chern indices in figure (\ref{fig:ellipses}).

%fig7:ellipses

%==================================================================
\subsection{The degeneracy lines are ellipses.} \label{sec:degeneracies}

We are now ready to present the final result.
Consider a generic two-dimensional submanifold  of the manifold of classical Hamiltonian $H_{\gamma}$ ($\gamma$ are coordinates on this manifold, and can be viewed as external parameters of the Hamiltonian). 
We have obtained that in this space, degeneracy lines are two ellipses (one for each pair of levels $(1,2)$ and $(2,3)$).

Note that it make sense to state that the shape of the curves are ellipse even if the $H_{\gamma}$-space is a manifold. This is because the ellipse are exponentially small, and thus live in the tangent space.

More precisely, there exist a coordinate system  $(\gamma_1,\gamma_2)$ (i.e. a special parametrization of the Hamiltonian) for which the two ellipses are given by Eq. (\ref{eq:ellipses}).
The width of the ellipses are given by the coefficients $K_1,K_2$, whereas the position of the centers are given by coefficients $sY_1,sY_2$, $s=\pm1$.

It is obvious from Eq. (\ref{eq:scales}) that $K_1,K_2$ are exponentially small in the semi-classical limit. But this is true also for  $Y_1,Y_2$. This is shown at the beginning of the appendix (\ref{sec:parameters}).

We have then two different generic situations in the semi-classical limit:

\begin{enumerate}
\item  case:  the non-diagonal terms $Y_1,Y_2$ are exponentially small 
with respect to the diagonal terms $K_1,K_2$. 
The two ellipses are identical, see figure \ref{fig:casA} p. \pageref{fig:casA}.
At the origin, the three bands of energy will have Chern indices
of $(\mp 1,\pm 2, \mp 1)$. Outside the ellipse, the Chern indices are
$(0,0,0)$. Note that in the $(\gamma _1,\gamma _2)$ parameter space,
the ellipse are exponentially small and exponentially flat because $K_1$ is exponentially small compare to $K_2$ (or the inverse).

%fig8:casA

\begin{figure}[htbp]
\epsfig{file=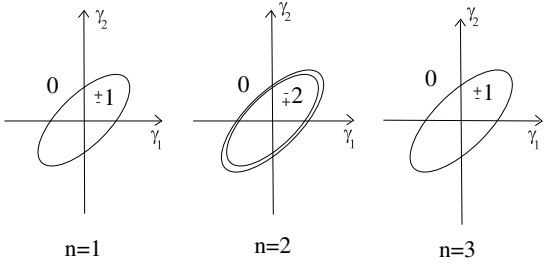, width=0.4\textwidth} 
\caption{Degeneracies lines and value of Chern indices in case 1.}
\label{fig:casA}
\end{figure}

\item case: the diagonal terms $K_1,K_2$ are exponentially small with respect to
the non-diagonal ones $Y_1,Y_2$. See figure \ref{fig:casB},
p. \pageref{fig:casB}.
At the scale $Y_i$ of the distance between the two center, the degeneracy lines collapse to 2 single points as $\hbar $ tends
to 0. Individually these lines are
still elliptical with respect to their own scale $K_i$.

%fig9:casB
\begin{figure}[htbp]
\epsfig{file=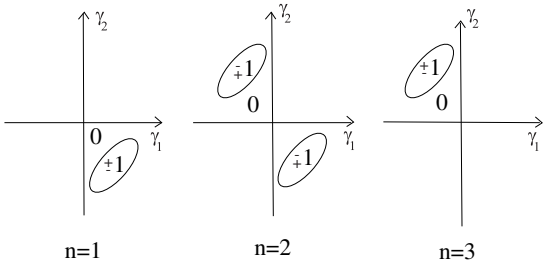,width=0.4\textwidth} 
\caption{Degeneracies lines and value of Chern indices in case 2.}
\label{fig:casB}
\end{figure}

\end{enumerate}

The non-generic intermediate case occurs when the diagonal terms $K_i$
and non-diagonal terms $Y_i$ are of the same order.
Hence the degeneracies are two ellipses ($s=\pm 1$) with symmetric centers. These two ellipses may intersect. Corrections to the leading behaviour in the semi-classical limit can modify the shape and sizes of the ellipses, see the numerical section.

%\begin{figure}[htbp]
%\epsfig{file=degen3_10b.eps,width=0.8\textwidth} 
%\caption{Degeneracies lines and value of Chern indices in case 3.}
%\label{fig:casC}
%\end{figure}

We now discuss the special  translation-symmetric case corresponding to the first calculation of Chern indices done by Thouless and al. \cite{iqhe2}. For example, take $H(q,p)$ from Eq. (\ref{eq:hamnum}) with $b=\gamma_1= \gamma_2=0$. See also remark \ref{rem:quantization}.
We construct $|2>={\hat T}_{P/3} |1>$ and $|3>={\hat T}_{P/3} |2>$
by translation by $P/3$ in the momentum direction. Applying the relation
$T_{P/3}T_Q=e^{2i\pi N/3}T_Q T_{P/3}$ we get:
\begin{equation} \label{eq:trimsym}
\varphi _{2}=\varphi _{1}/2=- \varphi _{1}= - \frac{2\pi N}{3} 
\end{equation}
Since the non-diagonal $m_{ij}$ are equal, so we are in the first case (double ellipse centered at the origin of  figure \ref{fig:casA}). Assuming that $3$ doesn't divide $N$ then we get $K_1=K_2$ and the precise shape of the ellipse:
\[ \gamma_1=\frac{ K}{\hbar} \cos(t), \quad \gamma_2= \frac{K}{\hbar} \cos(t \pm \frac{2\pi }{3})  \] 
We conclude that we get an ellipse
of axes $\gamma_1=\pm \gamma_2$ and excentricity $\sqrt{2/3}$.

\begin{rem} \label{rem:clusters}
We want to stress that all the results obtained in this section are valid
modulo error terms of order $O(r^2)$, where
$r=O(h^\infty)$ is the greatest tunnelling interaction between two wells. If one
of the previous contributions $m_{ij}$ between two other wells is smaller than $r^2$, it is not significant. For example, if we consider an Hamiltonian where wells
2 and 3 are very closed compared to well 1, our method must be modified, although we think that our results Eq. (\ref{eq:ellipses}) are still qualitatively correct.
In this situation, a more convenient approach is to first treat
the cluster of wells 2 and 3 and finally consider the interaction between
the cluster and well 1. By this approach, we obtain results similar to those described
in \cite{fred1} for the 2-wells problem.
\end{rem}

%==================================================================
\section{Numerical illustration}
More precise informations on how to perform the numerical
simulation, and a  software will be available on the web address: 
{\tt http://www-fourier.ujf-grenoble.fr/}\~{}{\tt  parisse}
or {\tt http://lpm2c.polycnrs-gre.fr/}\~{}{\tt faure}

We take the following Hamiltonian on the torus, parametrized by $\gamma_1$ and $\gamma_2$:

\begin{eqnarray}
\label{eq:hamnum}
 H(q,p)=&\cos (2 \pi \frac{q}{Q}) + \cos (6 \pi \frac{p}{P}) \\
        & + 4 b \sin(2 \pi \frac{q}{Q}) \cos(2 \pi \frac{p}{P}) \nonumber \\
        & + ((\gamma_1 - \gamma_2) \cos ( 2 \pi \frac{p}{P}) + ((-\gamma_1 - \gamma_2) \sin ( 2 \pi \frac{p}{P}) \nonumber
\end{eqnarray}

\begin{figure}[htbp]
\epsfig{file=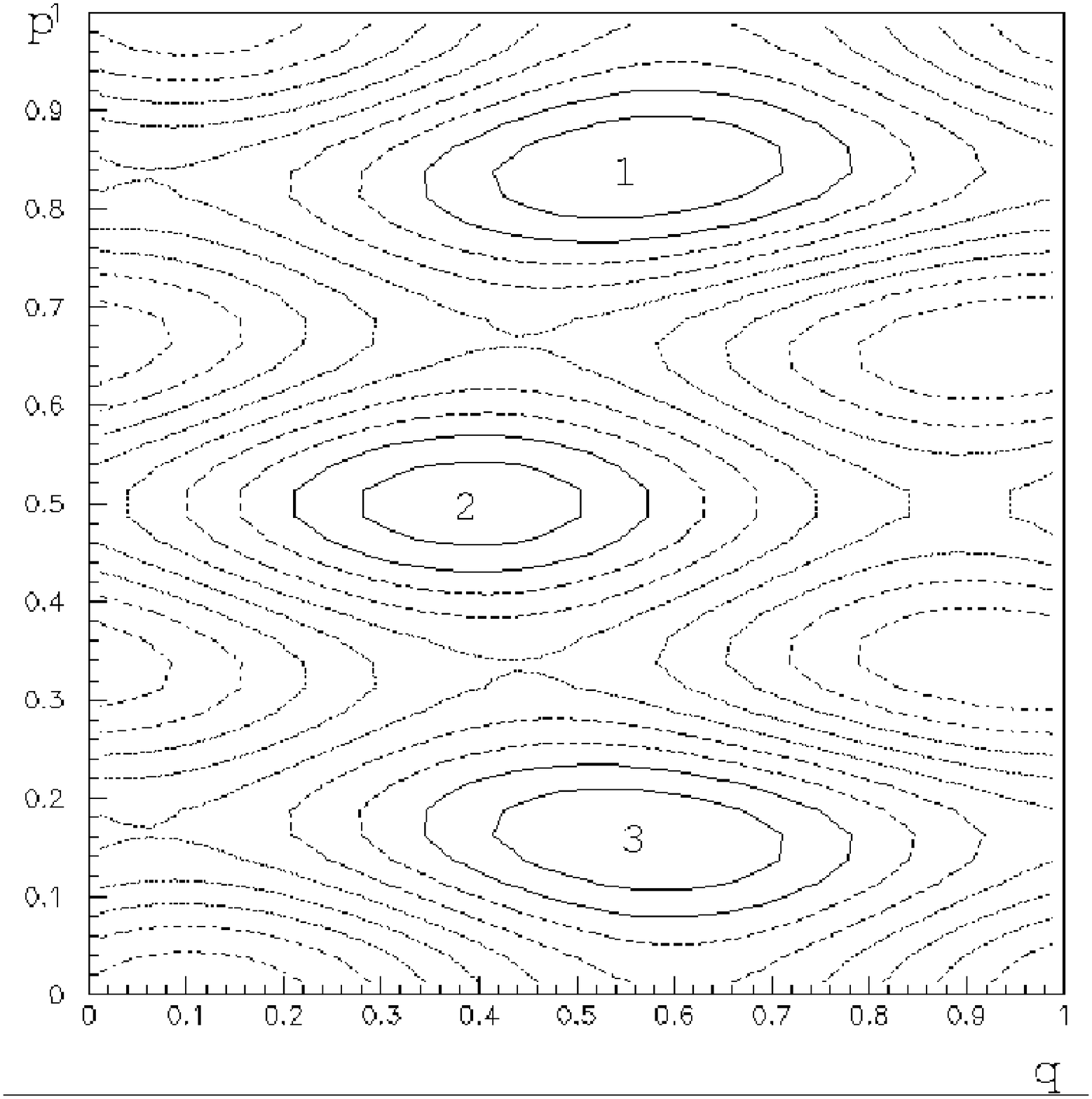,width=0.3\textwidth} 
\caption{Contour levels of Hamiltonian (\ref{eq:hamnum}), with $b=0.2$ and $\gamma_1=-\gamma_2=-0.1$}
\label{fig:hamnum}
\end{figure}

The two terms in the first line are the main part of the Hamiltonian. They create three symmetric wells in the $p$ direction in the lower part of the energy spectrum, see figure (\ref{fig:hamnum}). The second line breaks this symmetry, by moving the second well on the left when $b$ increases. $b$ can be seen as a perturbation from the translational symmetric case. The coefficients $\gamma_1$ and $\gamma_2$  in the third line change the depth of wells  1 and 3 respectively. We therefore expect them to act directely on
$( \tilde{E}_1-\tilde{E}_2)$ and $( \tilde{E}_3-\tilde{E}_2 )$ respectively, such that (\ref{eq:condition}) will be satisfied.
In figure  (\ref{fig:hamnum}) it seems clear that the main tunnelling interactions between the wells are those sketched in figure (\ref{fig:interaction}) as assumed for the calculations of the previous section. \\

%fig10:hamnum

\begin{figure}[htbp]
\epsfig{file=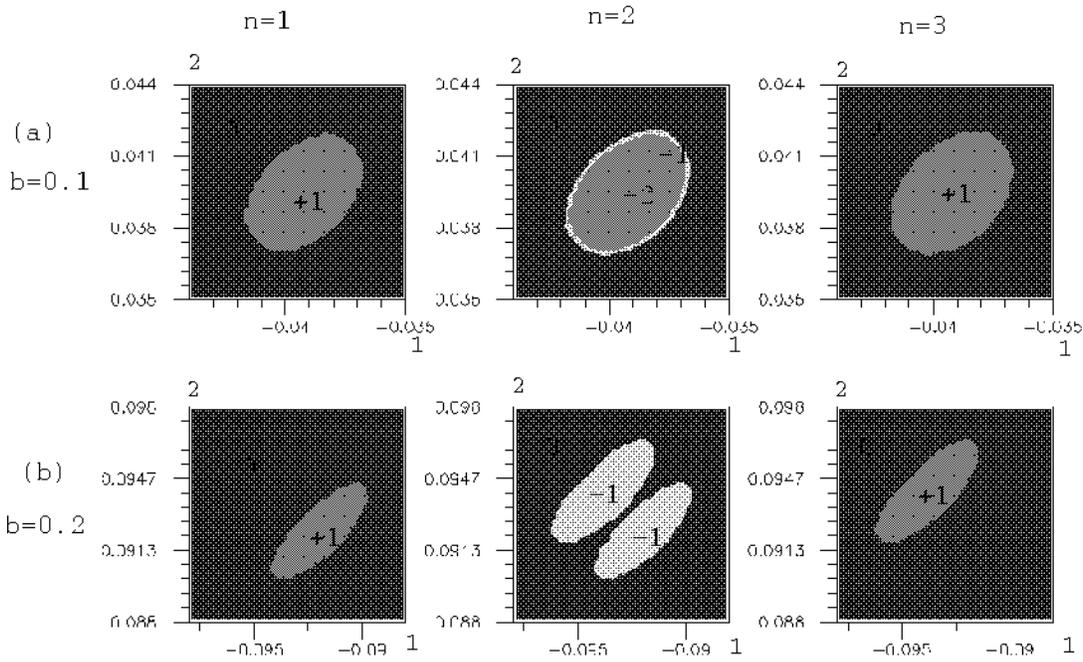, width=0.8\textwidth} 
\caption{Chern indices for the first three bands $n=1,2,3$, for $N=17$. 
  Case (a): The strength of the perturbation is $b=0.1$.
 Case (b): The strength of the perturbation is $b=0.2$.}
\label{fig:n17}
\end{figure}

For the numerical calculations, the Hamiltonian (\ref{eq:hamnum}) has been diagonalized and formula (\ref{e:chern}) has been used to obtain the Chern indices $C_n$. Maps of the Chern indices have been numerically obtained by varying parameters  $\gamma_1$ and $\gamma_2$.\\
For $N=17$ the values of the Chern indices for the first three energy bands $n=1,2,3$ are shown in figure \ref{fig:n17}. 

%fig11:n17

\begin{figure}[htbp]
\epsfig{file=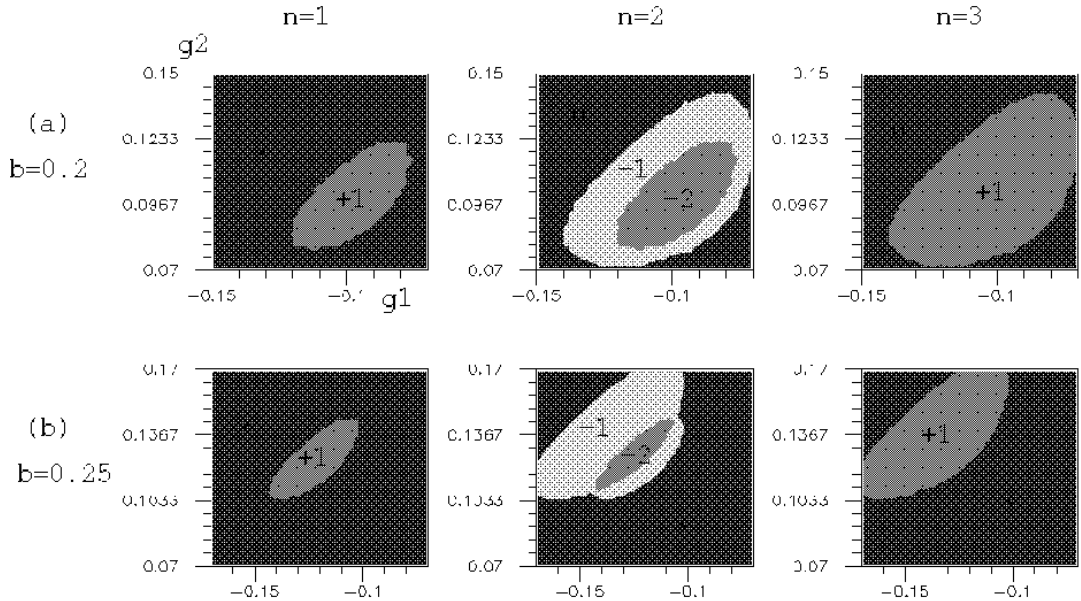, width=0.8\textwidth} 
\caption{Chern indices for the first three bands $n=1,2,3$, for $N=11$. 
  Case (a): The strength of the perturbation is $b=0.2$.
 Case (b): The strength of the perturbation is $b=0.25$.}
\label{fig:n11}
\end{figure}

 In case (a), the strength of the perturbation is $b=0.1$. The two ellipses are degenerated. This is the generic situation depicted in figure (\ref{fig:casA}), and  the same generic situation as for the translation-symmetric case with $b=0$, treated in the previous section. This means that the parameter $b=0.1$ is small enough to stay in the same generic ensemble. \\
In case (b), the strength of the perturbation is $b=0.2$. The two ellipses are separated, as in the second generic situation, sketched in figure  (\ref{fig:casB}). \\
In case (a) and (b), the two ellipses have the same shape, in agreement with  the semi-classical results of the previous section.\\
Figure \ref{fig:n11} shows results for a lower value of $N$, $N=11$, and for strengths $b=0.2$ and $b=0.25$. Here, the generic situations expected in the semi-classical limit   $N\rightarrow +\infty$ are less visible. There are still elliptical curves (from a topological point of view), but their position and shapes are affected by non-leading corrections with respect to  $N\rightarrow +\infty$.

%fig12:n11

%==================================================================
\section{Conclusion}
In this paper, we have shown in which conditions a non-zero Quantum Hall conductivity can occur in the framework of the Harper model, and for the tunnelling between three trajectories in a periodic cell. In this framework, the Hall conductivity is proportional to the topological Chern index. These results were derived semi-classically and give at the same time a description of degeneracies in the spectrum.

Currently, one tries to observe experimental signatures of the Harper spectrum (Landau level substructures) in lateral superlattices with periods of about $100 nm$ on GaAs-AlGaAs heterojunctions \cite{gudmundsson,schlosser}. Our  results could therefore have some experimental importance in the future. 

The main mathematical gap in this work is that we do not have results
about microlocal tunnelling effect (like those obtained by Helffer
and Sj\"ostrand \cite{HS} in the Schr\"odinger case). The main ideas to solve
these difficulties are currently majorations of the
wave function (A.Martinez \cite{martinez94}),
complex paths 
(estimations of the tunnelling effect by Wilkinson \cite{harper8,harper4}),
or normal forms (S.Nakamura \cite{nakamura95}),
but the authors are not aware of a general rigorous result.

Some other related problems could be investigated in the future like:
\begin{itemize}
\item numerical localization in the $\gamma $-space of the ellipsis
(this requires numeric estimations of the tunnelling effect),
\item generic interaction between $N$ bands ($N$-wells tunnelling effect)
by a recursive clustering approach,
\item the tunnelling effect between non-contractible classical trajectories
\end{itemize}

\vspace{0.3cm}
\begin{center}
{\Large \bf Acknowledgments}
\end{center}

We would like to thank P. Leboeuf for giving the initial impulse of this
work and Y. Colin de Verdi\`ere for his constant interest, his patient
relecture of this work and the subsequent remarks he made.

%\newpage

\appendix
%==================================================================
\section{Eigenvalue degeneracies in 3$\times $3 hermitian matrices.} 
\label{sec:degen3}

In the text, we use the following result of linear algebra that we
have not found in the literature:
\begin{theorem}
Let $A$ be a $3\times 3$ Hermitian matrix: 
\[
A=\left( 
\begin{array}{ccc}
a & \delta & \gamma \\ 
\bar \delta & b & \varepsilon \\ 
\bar \gamma & \bar \varepsilon & c
\end{array}
\right) 
\]
where $a,b,c$ are real numbers and $\gamma ,\delta ,\varepsilon $ are non-zero complex numbers. 

Then $A$ has a eigenvalue $\lambda $ with multiplicity 
at least $2$ if and only if: 
\[
a-\frac{\bar \delta \gamma }\varepsilon =b-\frac{\delta \varepsilon }\gamma
=c-\frac{\gamma \bar \varepsilon }\delta =\lambda \in R 
\]
(this is a system of $3$ real equations).
The third eigenvalue is $\lambda=tr(A)-2 \lambda$ and the spectrum is $(\lambda, \lambda,\lambda ')$ with $\lambda ' >\lambda$ iff $\Re(\delta \varepsilon \bar \gamma)>0$.
\end{theorem}

\emph{Proof}\\
Let us recall that a hermitian matrix has real eigenvalues and is
diagonalizable. Then $A$ has an eigenvalue $\lambda $ with multiplicity 2 or 3
if and only if its minimal polynomial has degree $2$ or $1.$ The matrix $A^2$
is then a linear combination of $I$ and $A$. We have:
\[
A^2=\left( 
\begin{array}{ccc}
a^2+\left| \delta \right| ^2+\left| \gamma \right| ^2 & (a+b)\delta +\gamma
\bar \varepsilon & (a+c)\gamma +\delta \varepsilon \\ 
\ast & b^2+\left| \varepsilon \right| ^2+\left| \delta \right| ^2 & 
(b+c)\varepsilon +\bar \delta \gamma \\ 
\ast & * & \left| \gamma \right| ^2+\left| \varepsilon \right| ^2+c^2
\end{array}
\right) 
\]
where the stars denote the conjugate of the symmetric coefficients.

The other eigenvalue is $\lambda'=tr(A)-2\lambda $, where $tr(A)$ 
denotes the trace of the
matrix $A$. Therefore the minimal polynomial is:
\[
P(X)=(X-\lambda )(X-\mbox{tr}(A)+2\lambda )=
X^2+(\lambda -\mbox{tr}(A))X-(2\lambda
-\mbox{tr}(A))\lambda 
\]
hence: 
\[
A^2=(\mbox{tr}(A)-\lambda )A+\lambda (2\lambda -\mbox{tr}(A)) 
\]
Eventually we get the system: 
\begin{equation}
\left\{ 
\begin{array}{rcrcr}
a^2+|\delta |^2+|\gamma |^2 & = & (\mbox{tr}(A)-\lambda )a & + & \lambda
(2\lambda -{tr}\,A) \\ 
b^2+|\delta |^2+|\varepsilon |^2 & = & (\mbox{tr}(A)-\lambda )b & + & \lambda
(2\lambda -{tr}\,A) \\ 
a^2+|\varepsilon |^2+|\gamma |^2 & = & (\mbox{tr}(A)-\lambda )c & + & \lambda
(2\lambda -{tr}\,A) \\ 
(a+b)\delta +\gamma \overline{\varepsilon } & = & (\mbox{tr}(A)-\lambda )\delta & 
&  \\ 
(a+c)\gamma +\delta \varepsilon & = & (\mbox{tr}(A)-\lambda )\gamma &  &  \\ 
(b+c)\varepsilon +\overline{\delta }\gamma & = & (\mbox{tr}(A)-\lambda
)\varepsilon &  & 
\end{array}
\right.  \label{e:syst}
\end{equation}
But $\mbox{tr}(A)=a+b+c$, so the last three equations simplify to: 
\[
\left\{ 
\begin{array}{rcr}
\gamma \overline{\varepsilon } & = & (c-\lambda )\delta \\ 
\delta \varepsilon & = & (b-\lambda )\gamma \\ 
\overline{\delta }\gamma & = & (a-\lambda )\varepsilon
\end{array}
\right. 
\]
When $\delta $, $\gamma $, and $\varepsilon $ are non-zero, we verify that
these three last equations imply the first three of system Eq. (\ref{e:syst}).
Indeed by multiplying two with two, we obtain: 
\[
\left\{ 
\begin{array}{rcr}
|\varepsilon |^2 & = & (b-\lambda )(c-\lambda ) \\ 
|\delta |^2 & = & (a-\lambda )(b-\lambda ) \\ 
|\gamma |^2 & = & (a-\lambda )(c-\lambda )
\end{array}
\right. . 
\]

 The spectrum is $(\lambda, \lambda,\lambda ')$ iff  $3 \lambda <tr(A)$. This gives $\Re(\delta \varepsilon \bar \gamma)>0$.

%==================================================================
%fred nov 97
\section{Influence of the external parameters} \label{sec:parameters}
In this appendix, we show that generically, only two external parameters  $\gamma_1,\gamma_2$ control  degeneracies and the Chern indices, as in Eq. (\ref{e:paramm}). 
  
From the generic cyclic assumption made in section \ref{sec:matrice33}, we have the property:
\[ \forall i \neq j, i \neq k, j\neq k, \quad 
m_{ij} m_{jk} \mbox{ is } O(h^\infty )\mbox{ with respect to }
m_{ik} \]
In the analogous Schr\"odinger situation, this means that
the triangular inequality is strict for the Agmon distance.
\begin{rem} \label{rem:generic}
For the Agmon distance, the converse assumption is also generic
(wells may be ``shaded'' by other wells). It is excluded here, because
assuming that a well is shaded implies that the cycle of trajectories
is contractible.
\end{rem}
For example, if $H$ is symmetric under translation ${\hat T}_{P/3}$, we see that:
\[ m_{ij} m_{jk}/m_{ik} = m_{ij} = O( r)=O(h^\infty ) \]
Then we get easily that $\tilde{E}_i-\tilde{E}_j$ must be $O(h^\infty )$
in Eq. (\ref{eq:newdeg}).

More precisely, we want to look at the $\gamma $-dependency of the
two equations of (\ref{eq:newdeg}). From the first equation, we define
\[ f_1(\gamma ,\theta _1)=
\frac{1}{\hbar } \left(
(\tilde{E}_1 - \tilde{E}_3) - \left[
s m_{13} \left( \frac{ m_{12} }{ m_{23} } - \frac{ m_{23} }{ m_{12} }  \right)
+ m_{33} \cos(\varphi _{33}-\theta_1 )
- m_{11} \cos(\varphi _{11}-\theta_1 )
\right]  \right)    \]
and we define $f_2$ similarly using the second equation of (\ref{eq:newdeg})
so that (\ref{eq:newdeg}) becomes 
\[f(\gamma , \theta _1)=0 \]
where 
\[ f: \left\{ \begin{array}{rcl}
 I\!\!R ^p\times [0,2\pi ]  &\rightarrow &I\!\!R ^2  \\
(\gamma ,\theta ) & \rightarrow & (f_1,f_2)
\end{array} \right. \]
We see easily that:
\[ d  f= \frac{1}{\hbar } 
d( \tilde{E}_1-\tilde{E}_3, \tilde{E}_2 -\tilde{E}_3 ) + O(h^\infty ) \]
If we want to have the simplest possible parametrization of the degeneracies, 
we have to find 2 parameters such that:
\[ J= 
\partial_{\mbox{2 parameters}} ( \frac{1}{\hbar } (\tilde{E}_1-\tilde{E}_3), 
\frac{1}{\hbar } (\tilde{E}_2-\tilde{E}_3) ) \]
is invertible. Hence, we can not take $\theta _1$ as a parameter,
and we must take two $\gamma $ parameters. This means that the degeneracy
will not be described by a point on a 1 parameter-$\gamma $ line but
by a line on a 2 parameters-$\gamma $ plane.
We now assume that the parameters $\gamma _1$ and $\gamma _2$ satisfy
the invertibilty hypothesis, and more precisely that there exists a constant
$C$ such that:
\begin{equation}
\label{eq:condition}
|J| \leq C ,
\quad
|J^{-1}| \leq C, \quad \mbox{with }
J=\partial_{\gamma _1,\gamma _2} 
( \frac{1}{\hbar } (\tilde{E}_1-\tilde{E}_3), 
\frac{1}{\hbar } (\tilde{E}_2-\tilde{E}_3) )
\end{equation}
for $\gamma $ in an neighboorhood of a point $\gamma ^0$. 
In addition, if we suppose that $\gamma^0 $ is an approximate 
solution of $f=0$: 
\[ f(\gamma^0,\theta _1 )=O(h^\infty ).\]
we can apply the implicit function theorem 
with parameter $\hbar ,\theta _1, \gamma _3,..., \gamma _p$,
for $\hbar\in ]0,h_0] $ with $h_0$ sufficiently small such that
the rest terms $O(h_0^\infty )$ are smaller than the constants $C$). 
We get the existence of an $\hbar $-independent neighboorhoud 
$U$ of $((\gamma^0)_1,(\gamma ^0)_2)$ 
such that for every fixed 
$(\gamma _3,...,\gamma _p)$ in a neighboorhoud of 
$ ( (\gamma ^0)_3, ..., (\gamma ^0)_p)$
and for every fixed $\theta _1$, the function
$f(.,\gamma _3,...,\gamma _p,\theta _1)$ of $(\gamma _1,\gamma _2)$
is bijective from $U$ to $f(U)$. 
Since $f(\gamma ^0,\theta _1)=O(h^\infty )$, 
there exists a unique point $\gamma (\theta _1,\gamma _3,...,\gamma _p)$ in $U$
such that $f(\gamma(\theta _1,\gamma _3,...,\gamma _p),\theta _1 )=0$ 
and $|\gamma(\theta _1,\gamma _3,...,\gamma _p) -\gamma ^0|=O(h^\infty )$.

If we fix $(\gamma _3, ...., \gamma _p)$ and let $\theta _1$ move, the
projection of the point $\gamma (\theta _1,\gamma _3,...,\gamma _p)$ 
in the $(\gamma _1,\gamma _2)$ space describes a curve of size $O(h^\infty )$.
At that scale, all classical quantities 
(like tunneling interactions) are constant 
(with a relative error of order $O(h^\infty )$).

%======================================================================
\section{More general degeneracy curves} \label{sec:lissajou}

In this appendix, we solve the general problem raised in section \ref{sec:matrice33}, i.e. we look for the expression of the degeneracy curves in the space $X_2,X_3$ for any value of 
$\vec{n}_{ij}, i,j=1,2,3$.

Let us put
\[
\vec{N}=\vec{n}_{12}+\vec{n}_{23}+\vec{n}_{31}
\]
The first equation  of Eq. (\ref{e:degen}) gives:
\begin{equation}
\Theta =\varphi -\vec{N}\vec{\theta} \equiv 0\qquad [\pi ]  \label{e:cond1}
\end{equation}
with $\varphi= \varphi _{12}+\varphi _{23}+\varphi
_{31}$.
We put  $s= \exp (i \Theta )$.
We suppose (see paragraph (\ref{sec:matrice33}) ) that 
\[
\vec{N} \ne \vec{0}
\]
There is an unique decomposition of $\vec{N}$ as $\vec{N}=p \vec{M}$ where $p \in N^*$ is the greatest common divisor of $(N_1,N_2)$, so $(M_1,M_2)$ are relatively prime. Then Eq. (\ref{e:cond1}) gives 
\begin{equation}  \label{e:constr}
\vec{M} \vec{\theta}=(\varphi+k \pi)/p, \qquad k=1,...,2p
\end{equation}
$k$ even (respect. odd) corresponds to $s=+1$ and degeneracy between $n=1,2$ (respect.  $s=-1$ and degeneracy between $n=2,3$).

\begin{figure}[htbp]
\epsfig{file=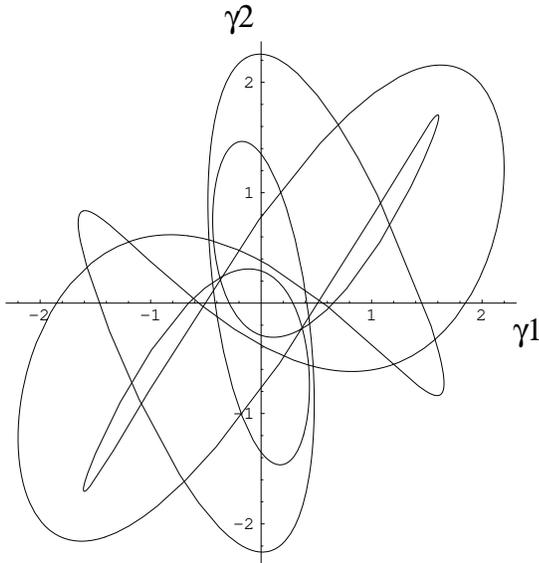, width=0.4\textwidth} 
\caption{Parametrized curve $\gamma_1(t)=1.2 \cos(5t+1)-\cos(3t)$,  $\gamma_2(t)=1.3 \cos(7t+0.3)-\cos(3t)$ as an example of the solution Eq.( \ref{e:solution}) for a general degeneracy curve.}
\label{fig:curves}
\end{figure}

The last two  equations  of Eq. (\ref{e:degen}) give the curves equations as:
\begin{eqnarray} 
\gamma_1= \hbar( -sY_1 + m_{11} \cos(\varphi _{11}-\vec{n_{11}} \vec{\theta} )
- m_{33} \cos(\varphi _{33}-\vec{n_{33}} \vec{\theta}  ))
\\
\gamma_2=  \hbar(-sY_2 + m_{22} \cos(\varphi _{22}-\vec{n_{22}} \vec{\theta} )
- m_{33} \cos(\varphi _{33}-\vec{n_{33}} \vec{\theta}  ))
\end{eqnarray}
with $\gamma_1, Y_i$ defined in paragraph (\ref{sec:matrice33}) and  $\vec{\theta}$ is allowed to vary but constrained by Eq. (\ref{e:constr}).
Because $\vec{M}=(M_1,M_2)$ are relatively prime, there exists $\vec{n}_0$ from B\'ezout 's theorem such that $\vec{M}$ and $\vec{n}_0$ form a basis of  the $Z^2$ lattice, that is $\det [\vec{M}, \vec{n}_0 ]=1$.
In this basis,
we can decompose:  $\vec{n}_{ii}=b_i \vec{M} + c_i \vec{n}_0$, with $b_i,c_i \in Z$ for $i=1,2,3$. This gives 
\[ \varphi _{ii}-\vec{n_{ii}} \vec{\theta}= \tau_{i,k} -c_i t \]
 with 
\[
\tau_{i,k} = \varphi _{ii} - b_i \varphi /p - b_i k 2\pi /p , \quad k=1,..,2p  \quad i=1,2,3
\]
and a free parameter $t=\vec{n}_0 \vec{\theta} \in I\!\!R$.The final expression for the degeneracy curves is:
\begin{eqnarray}  \label{e:solution}
\gamma_1(t)=  \hbar(-sY_1 + m_{11} \cos(\tau_{1,k} -c_1 t )
- m_{33} \cos(\tau_{3,k} -c_3 t ))
\\
\gamma_2(t)= \hbar( -sY_2 + m_{22} \cos(\tau_{2,k} -c_2 t )
- m_{33}  \cos(\tau_{3,k} -c_3 t ))
\end{eqnarray}
with $k=1,..,2p$. This gives $2p$ curves. They can be quite complicated in general. See an example in figure (\ref{fig:curves}). These curves are degenerate if all $c_i=0$, this means that every $\vec{n}_{ii}$ is proportionnal to $\vec{N}$.
As pointed out in  paragraph (\ref{sec:matrice33}), we think that the 
only possible situation in the Harper model considered in this paper is the 
ellipse curve, Eq.(\ref{eq:ellipses}), for which $p=1$, $c_1=c_2=c_3$.

%fig13:curves

%\newpage

\bibliography{preprint}
\end{document}